# Supercritical water: percolation transitions and a colloidal mesophase*

**Leslie V. Woodcock**

Department of Physics, University of Algarve, Faro 8000, Portugal

ABSTRACT

A revised phase diagram for water shows three distinct fluid phases. There is no continuity of liquid and gas, and no 'critical point' on Gibbs' density surface as hypothesized by van der Waals. A liquid state, water, spans all temperatures. A supercritical colloidal mesophase bounded by percolation transition loci separates supercritical liquid water and gas-phase steam. The liquid water phase is bounded by a percolation transition (PA) of available volume, whereas the gas phase steam is bounded by transition loci when a bonded molecular cluster suddenly percolates. At the percolation loci there is no barrier to nucleation of water to steam (PA) or steam to water (PB). Below the critical temperature the percolation loci become the metastable spinodals in the two-phase coexistence region. A critical divide is defined by the interception of PA and PB the p-T plane. Critical parameters are obtainable from slopes and intercepts of pressure-density supercritical isotherms within the mesophase. The supercritical mesophase is a fourth equilibrium state besides ice, water and steam. The thermodynamic state function rigidity $(dp/d\rho)_T$, which phenomenologically defines a distinction between liquid and gas, shows a remarkable symmetry due to an equivalence in number density fluctuations, arising from available holes and molecular clusters, in liquid and gas respectively. Every state of steam will have a corresponding complimentary state of water at the same temperature.

## 1 INTRODUCTION

The thermodynamic diagram of water phases is described in a recent educational review [1]. Theoretical and computational advances, however, have done little to change the generally accepted phase boundaries of water and steam over many decades [2]. A wide variety of water phase diagrams in various projections of Gibbs surfaces for state functions of temperature (T) and pressure (p) can be viewed on comprehensive websites [3]. They show no boundaries between the phases of water and steam as they merge above the critical temperature. Some show *ad hoc* dividing lines using the critical isotherm, others the critical isobar, to bound "water" and "steam" from a supercritical "fluid", but these lines have no thermodynamic status.

The fluid phase diagram of water, and indeed all other pure fluids, is seen to be based upon the van der Waals hypothesis [4]. Briefly, for every fluid there is deemed to exist a critical temperature-pressure singularity at $T_c,p_c$ whereupon an equation-of-state for which the first two derivatives of pressure, i.e. with respect to changes in volume at constant T, go to zero. This has been the accepted description of liquid-gas criticality since van der Waals published his renowned thesis "On the continuity of liquid and gas". The hypothesis further implies that for all supercritical temperatures the thermodynamic state functions e.g. density $\rho(T,p)$ or Gibbs energy $G(T,p)$, from the gas at low density, steam, to the liquid at high density, water, are continuous in all their derivatives for supercritical temperatures across the whole density range.

Another hypothesis that has been fashionable amongst theoretical physicists for about 50 years is the concept of universality [5]. From Ising models, ferro-magnetic systems, spin glasses, to liquid-gas criticality, according to this community, all critical phenomena, obey the same universal physical description, in the vicinity of a "critical point", which can been explained using group renormalization theory [5]. An originator of this theory, K. Wilson, summarizes the phenomenology of liquid-gas criticality by reference to water and steam in one brief paragraph as follows [6]:

*"A critical point is a special example of a phase transition. Consider, e.g. the water-steam transition. Suppose that water and steam are placed under pressure always at the boiling temperature. At the critical point, the distinction between water and steam disappears, and the whole boiling phenomena vanishes. The principal distinction between water and steam is that they have different densities. As the pressure and temperature approach their critical values the difference in density between water and steam goes to zero".*

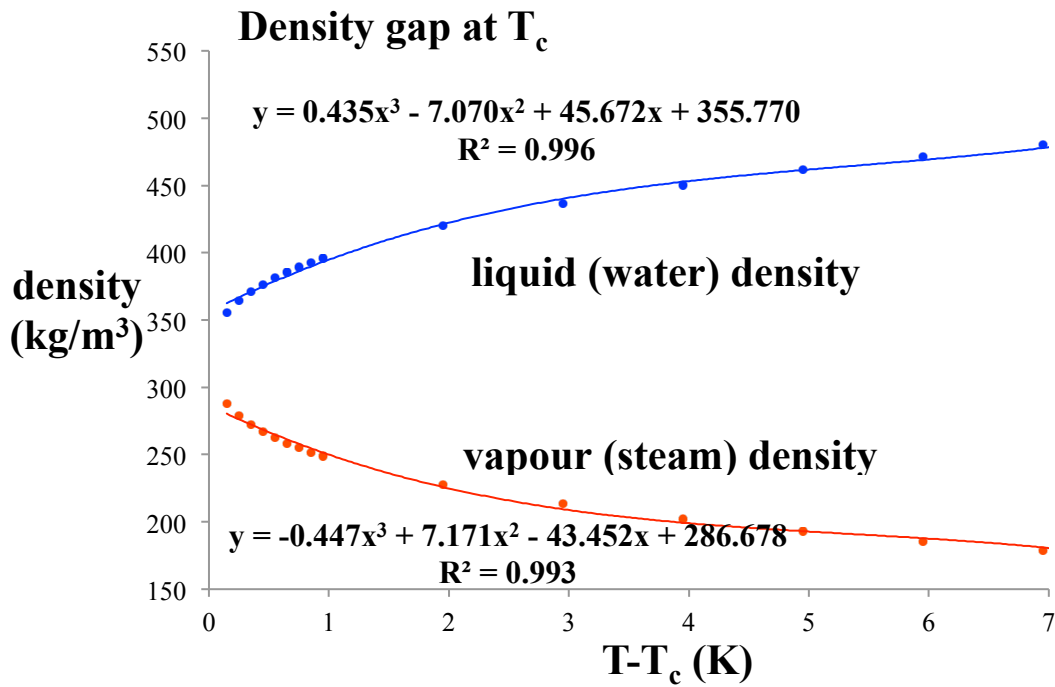

**Figure 1** Densities of saturated water and steam vapour in the vicinity of the critical temperature taken from IAWPS –IF97 Electronic Steam Tables (see reference [11]). The nearest data point to $T_c$ is at 373.8°C, i.e. within 0.1K of the critical temperature (373.9°C) from the same source; the critical density values are intercepts given by the EXCEL cubic trendlines with the constants representing the critical coexistence densities.

It is evident from modern 5-figure precision density data (Fig. 1), that Wilson's assertion does not accord with experiment. Recent investigations, moreover, of percolation transitions in model square-well fluids [7,8], also contradict this hypothesis. We have concluded that, at the critical temperature $T_c$, the density difference does not go to zero. Rather, we have found that a liquid-gas critical "point", as hypothesised by van der Waals and accepted by the physics community for 140 years, does not exist a such. Moreover, there is no universality of critical phenomena that spans the dimensions. In the ρ(T,p) surface there is a dividing line of uniform chemical potential connecting maximum and minimum coexisting gas and liquid densities, respectively, above which there exists a supercritical mesophase bounded by weak higher-order percolation transitions, and below which there is the familiar liquid-vapour coexistence region. Furthermore, and contrary to Wilson's statement above [6], for $T > T_c$ water and steam remain distinctly different phases, and are separated by a supercritical mesophase.

This new description of liquid-gas criticality has also been shown to describe all the experimental thermodynamic phenomenology of liquid and gaseous argon [9]. Water and steam exhibit essentially the same phenomenology as fluid phases of argon [5,6]. Moreover, for temperatures above the critical temperature the phases of water and steam exist with distinct liquid and gas properties, and are separated by a supercritical mesophase (formerly referred to as "steam II" [8]). The mesophase can be characterized as being macroscopically homogeneous, but microscopically heterogeneous colloidal-like mixture of gas-like molecules and molecular clusters, and liquid-like macroscopic clusters or droplets, all with the same Gibbs chemical potential. These phase boundaries appear as a weak higher-order thermodynamic phase transition, known as percolation transition loci, on the Gibbs state function f(T,p) surfaces [7-9]. They were unforeseen by Gibbs, but recently have been shown to play a role in determining the phase diagram alongside Gibbs phase rule [10].

## 2 PERCOLATION TRANSITIONS

Percolation transitions occur when structural properties of a molecular system, for example clusters of molecules in the gas phase, or holes in the liquid phase, on reaching a certain density suddenly coalesce from being of molecular dimension to become macroscopic, i.e. system spanning. This gives rise to higher-order thermodynamic phase transitions, in which there are discontinuities in second and/or higher derivatives of Gibbs chemical potential (μ) with respect to temperature or pressure, notably: isothermal compressibility $(d_2\mu/dp^2)_T$, heat capacity $(d_2\mu/dT^2)_p$ and thermal expansivity $(d_2\mu/dpdT)$ all of which undergo some degree of change at, or in the vicinity of, percolation transitions.

We begin therefore by taking a look at the behaviour of these three second-order thermodynamic state functions in the case of supercritical water (Fig. 2)

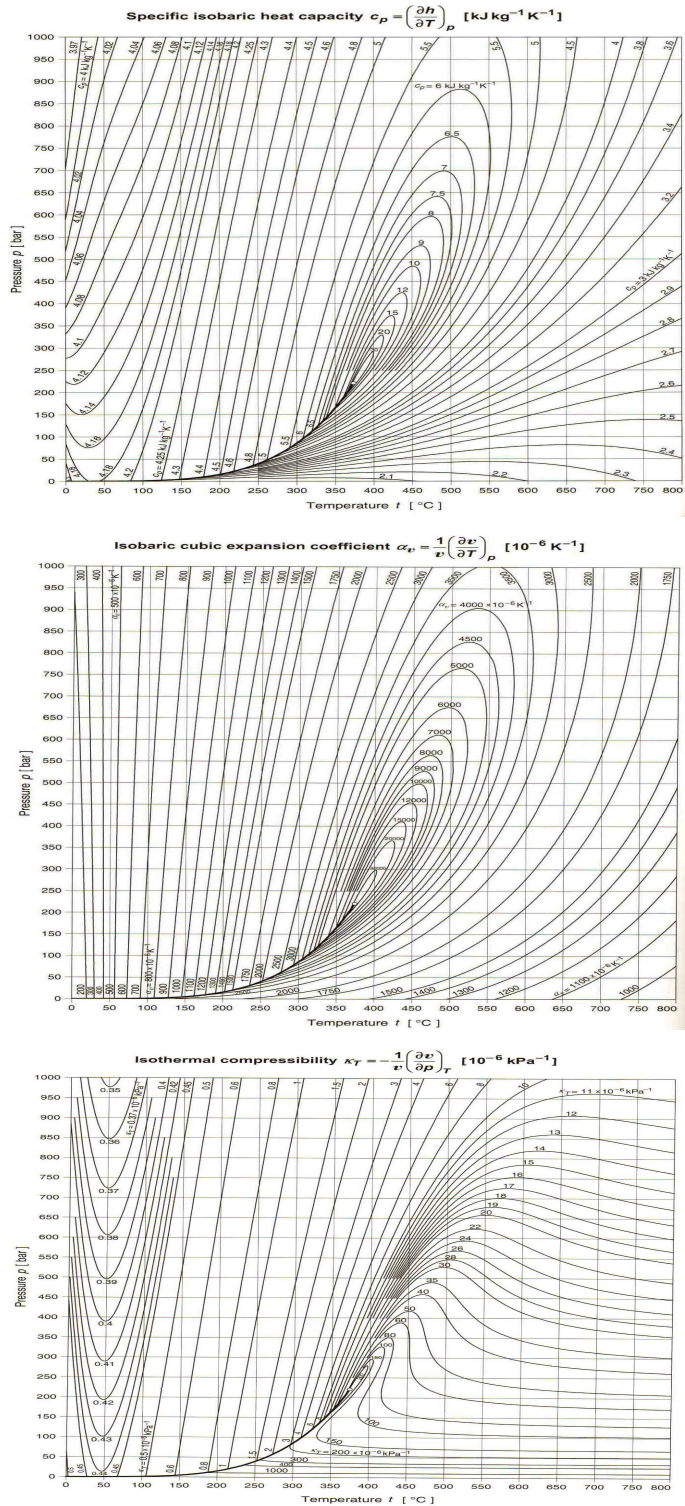

**Figure 2:** Isocontours for water and steam: (**a**) heat capacity ($C_p$) (**b**) thermal expansivity ($\alpha_p$) (**c**) isothermal compressibility ($\kappa_T$): from International Steam Tables [11].

The first observation from Fig. 2 is that all three derived properties exhibit a line of maximum values stemming from the coexistence line at $T_c$. The observation of these lines has a long history that goes back to J. D. Bernal [12]. They are characterized in the literature, but have remained somewhat of a mystery. Bernal actually named the line of maximum heat capacity ($C_p$), which can be clearly observed in Fig. 2a, as the "hypercritical line". It appears to be the extension of the coexistence line. The maximum in the thermal expansivity ($\alpha_p$) (Fig.2b) is shifted, very slightly to higher temperature from Bernal's hypercritical line; it is not quite the same line. Likewise, the line of maximum isothermal compressibility Fig. 2c, which appears first to have been identified by Widom [13] is also shifted even further to higher temperatures. Interestingly, these three supercritical loci have been rediscovered recently in molecular simulations of Lennard-Jones fluids [14]. Associated lines of maxima in various dynamical properties have also been reported [15], a Russian name has even been suggested for another such line discovered recently [16].

The second observation that we make from Fig. 2 is that there is a clear difference in thermodynamic behavior between steam, at pressures below the coexistence - hypercritical line, and water above it. In the liquid phase, $C_p$, $\alpha_p$ and $\kappa_T$, all increase with temperature along an isobar, whereas in the gaseous state they all decrease with temperature. Thus there is a fundamental thermodynamic defining distinction between liquid and gas which extends to supercritical temperatures. Below, we find that defining property is the work required to change the density, i.e. an inverse compressibility, or rigidity $(dp/d\rho)_T$ where $\rho$ is the density. The rigidity of a gas decreases with density and increases for a liquid, i.e. $(d_2p/d\rho^2)_T$ is negative for a gas and positive for a liquid; its value is zero in the mesopphase. Lines of higher-order discontinuity in the rigidity can be characterized, not just for water, but all liquids, in terms of percolation loci that bound the supercritical mesophase separating gas from liquid.

Percolation transitions in model fluids of hard-spheres and square-well molecules are well defined by a characteristic distance, which unambiguously divides configuration space volume into 'sites' and 'holes'. The densities at which small clusters of occupied sites in the gas phase, or unoccupied holes in the liquid phase, first span the whole system are referred to as percolation transitions, and denoted by PB and PA, referring to 'bonded cluster' (sites) and 'available volume' (holes) respectively [7-9]. For model fluids with continuous pair potentials, and all real fluids, however, there is no such well-defined distance to define PA and PB. In the case of water, however, the steep hydrogen bond is fairly sharply defined for water clusters. At a given temperature, the total volume of any equilibrium configuration of water or steam can be divided in occupied sites, not accessible to an incoming molecule, and holes which would be accessible to an additional molecule. At these percolation transitions, thermodynamic state functions can change form due to sudden changes in the state-dependence of density and/or energy fluctuations. It is the change in $(dp/d\rho)_T$ that defines PA and PB phenomenologically.

In order to understand 'mysterious lines' [12-16] on the p(T) projection, we first look at the discontinuities in the Gibbs density surface which reflect the percolation loci PA and PB. These can be seen in Fig. 3, in which various supercritical isotherms from the International Steam Tables of Wagner and Krezschmar [11] are plotted and analysed for the three supercritical phases, as in a previous analysis of liquid argon [9] The bonded-cluster percolation transition (PB) occurs when atoms bonded together in molecular clusters, within a characteristic distance, e.g. close to the hydrogen-bond length, suddenly begin to be macroscopic. At a higher pressure and density, the available volume percolation transition (PA) occurs at a density below which the distribution of 'holes' in the liquid suddenly begin to permeate the volume or become macroscopic.

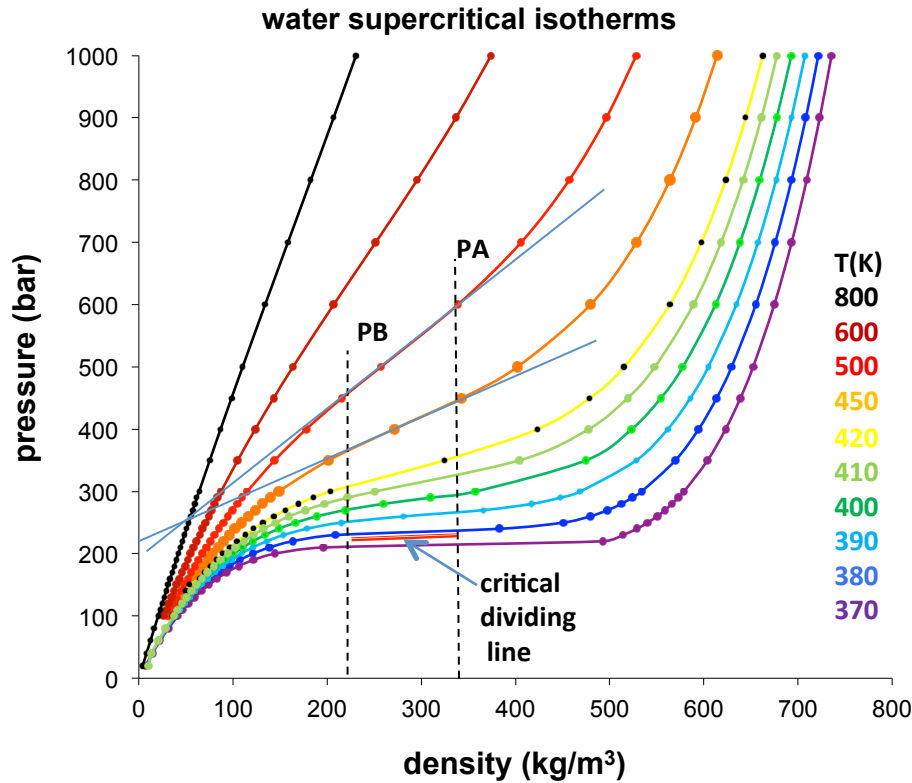

**Figure 3** Experimental data points for the p-ρ isotherms of supercritical water from the IAPWS-International Steam Tables [5]. The extended straight lines, fitted in the intermediate linear region for the two isotherms 450 and 500, have been superimposed to highlight the higher-order discontinuities associated with percolation transition loci PB and PA, approximately illustrated by vertical dashed lines at the critical coexisting densities in Fig.1.

The critical divide from Fig.1 at $T_c$ is shown in Fig. 3 from the experimental p-$\rho_T$ data [11]; the percolation loci must pass through these two points. The data points used to estimate the region of the mesophase, i.e. stemming from the two percolation transitions at $T_c$, are obtained foom the EXCEL trendline fits to the densities of water and steam in Fig.1  The lowest temperature in Fig.3, T=370, is subcritical ($T_c$ = 374°C). The percolation transitions define the limits of existence of the supercritical steam and water phases on the Gibbs density surface.

## 3  LIMITS OF GAS AND LIQUID

Both percolation transitions can be observed on all the supercritical isotherms as a weak discontinuity in the slope of the p(ρ) isotherms. For temperatures up to 800°C, all the isotherms show three distinctly different regions of behavior for the state function $p(\rho)_T$. In the steam phase $\rho < \rho_{PB}$ the rigidity $(dp/d\rho)_T$ decreases with density, in the mesophase $\rho_{PB} < \rho < \rho_{PA}$ $(dp/d\rho)_T$ is constant, and for water $\rho > \rho_{PA}$ and $(dp/d\rho)_T$ increases with density. The slope of the isotherm in the mesophase is zero at the critical temperature. This fundamental extensive state function, called "rigidity", is the work required to change the density reversibly, with the dimensions of a molar energy, hence equal to the change in Gibbs energy (G) with density

$$(dp/d\rho)_T = \rho(dG/d\rho)_T \qquad (1)$$

It will eventually explain all the mysterious maxima in Fig.2, the percolation loci in Fig.3, the nature of the supercritical mesophase, and it defines the distinction between steam and water or, more generally, between gas and liquid.

The reason for linearity of p(ρ) in the mesophase, can be understood if the percolation loci, although lacking a definition at the molecular level, are defined phenomenologically similar to subcritical spinodal lines. The limit of existence of the gas phase (PB) corresponds to the density ($\rho_{PB}$) at which the first system-spanning cluster of 'occupied sites' or stable liquid state 'nucleite' spontaneously appears in equilibrium configurations. The chemical potential of the small cluster species in the gas phase, becomes equal to the chemical potential of a liquid 'droplet' at a higher pressure. At this density the gas phase ceases to exist as such, and for densities above $\rho_{PB}$, i.e. in the mesophase, the fluid is essentially a homogeneous mixture of gaseous molecules, i.e. dimers, trimers and small clusters, etc. plus liquid-like macro clusters. There is a bimodal distribution of cluster sizes [9].

Likewise, as the liquid state density is reduced across PA, from above $\rho_{PA}$ to below $\rho_{PA}$, the 'unoccupied holes' spontaneously coalesce to form system-spanning 'holes' of gas which can contain single molecules, or small clusters, which will have the same chemical potential as the liquid, albeit at the lower gas phase pressure. Thus, in this region there are larger fluctuations in energy, pressure and density which explains the maxima loci of $C_p$, $\alpha_p$ and $\kappa_T$ in Fig. 2 and references [12 to 16].

The discontinuities in $(dp/d\rho)_T$, and linearity of $(dp/d\rho)_T$ in the mesophase region, seen in Fig. 3, has been simply explained by a linear combination rule. It is essentially a supercritical extension of the subcritical two-phase Lever rule. The structural properties of the mesophase in the vicinity of $T_c$ and PB will resemble a liquid-in-gas 'mist' with wide polydispersity of the micro-droplets, whereas in the vicinity of PA a colloid with water being the continuous phase, and steam dispersed rather like small bubbles as in a gas-in-liquid 'foam'. The different refractive index and size distribution of a dispersed phase give rise to a supercritical opalescence, or white appearance, known in colloid science as Tyndall scattering [17].

For temperatures below $T_c$, PA and PB become the metastable spinodals of liquid to gas, and gas to liquid, spontaneous nucleation loci respectively. Both above and below the critical divide at $T_c$, PA and PB represent the limits of existence of the liquid water and gaseous steam phases respectively.

Another feature of second or higher-order phase transitions, which is consistent with thermodynamic properties, is the non-existence of supercritical metastability. Whereas first-order phase transitions on Gibbs density surfaces are characterized by a possible metastable existence of either phase on both sides of the transition point of equal chemical potential, there can be no metastability across a second-order phase transition. The reason being since, in a higher-order transition, $(dp/d\rho)_T$ is the same for both phases at the point of the transition it follows from equation (1) that the phase with the higher rigidity would have a lower chemical potential, and hence be the more stable, on both sides of the transition on the Gibbs density surface. Since this is a thermodynamic impossibility, PA and PB loci are minima and maxima density limits for the existence of the liquid and gas phases, respectively.

## 4 CRITICAL PARAMETERS

Supercritical isotherms for experimental $\rho(p,T)$ data in the near-critical mesophase region can be used to obtain independent critical parameters without resort to experimental data in the two-phase region. In this section we perform the same analysis that resulted in the determination or revised critical properties of argon, but the present exercise is less accurate as there are fewer data points on each isotherm in the tables used [11].

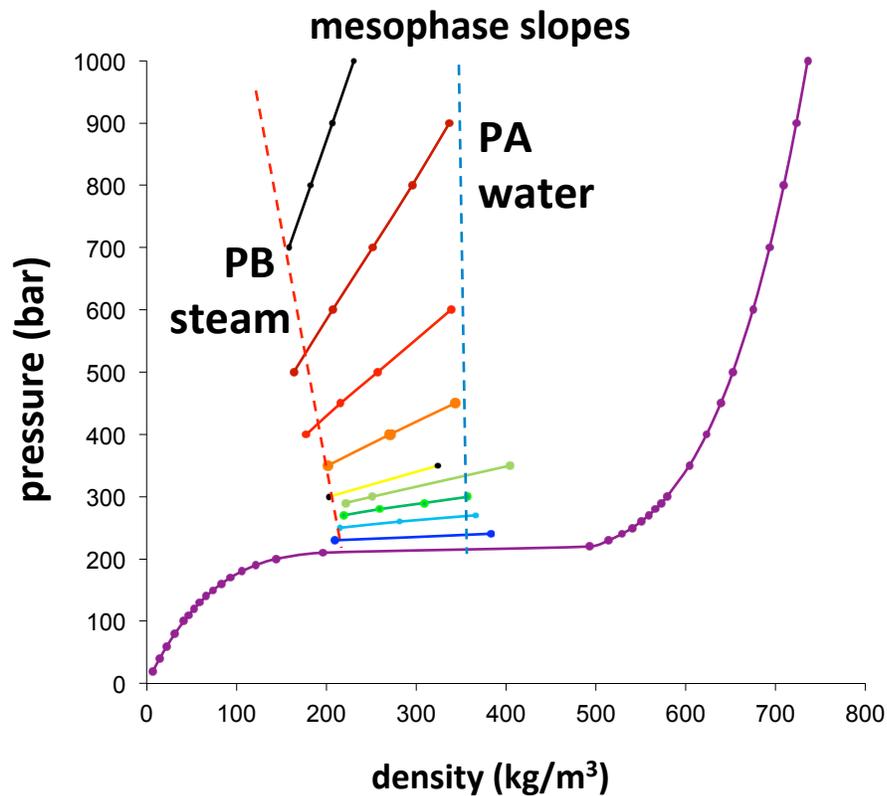

**Figure 4** Plots of the data points in the mesophase of nine supercritical isotherms taken numerically from the tables of Wagner and Kretzschmar [11] leading to the slopes and intercepts of equation (2) to determine critical parameters without the assumption of a "critical point"; the temperatures and colour coding are the same as in Fig.3.

**Table 1** Parameters in Eq. (2) for supercritical mesophase isotherms $p(\rho)_T$ ; the percolation loci are in units of density, kg.m$^{-3}$ and pressure in bar. The present critical parameters (bottom row in bold) correspond to rigidity slope $R_m= 0$

| isotherm (°C) | slope (J/kg) | intercept (bar) | density PB | pressure PB | density PA | pressure PA |
|---|---|---|---|---|---|---|
| 800 | 4.14579 | 43.28 | 162.2 | 715.8 | 230.6 | 999.5 |
| 600 | 2.25628 | 132.9 | 222.6 | 635.1 | 310.0 | 799.7 |
| 500 | 1.21997 | 186.6 | 241.3 | 480.9 | 349.5 | 572.8 |
| 450 | 0.72309 | 204.2 | 239.1 | 377.1 | 364.0 | 444.8 |
| 420 | 0.41230 | 216.2 | 242.9 | 316.4 | 373.4 | 355.2 |
| 410 | 0.32650 | 218.1 | 249.9 | 299.6 | 358.7 | 335.2 |
| 400 | 0.22003 | 222.2 | 254.2 | 278.1 | 357.7 | 300.9 |
| 390 | 0.13225 | 222.0 | 247.8 | 254.8 | 356.0 | 269.1 |
| 380 | 0.05756 | 217.9 | 238.1 | 231.7 | 370.2 | 239.3 |
| **374.7** | **0** | **221.2** | **234.3** | **221.2** | **370.9** | **221.2** |

Every supercritical isotherm, shown for example in Fig. 3, in the mesophase region, obeys the linear equation for the pressure with a high precision within the bounds of uncertainty. The linear equation

$$p_m = p_o + R_m \rho \qquad (2)$$

where $R_m$ is the constant isothermal rigidity $(dp/d\rho)_T$ within the mesophase. It is independent of density, and as shown in Fig.4, can be easily parameterized. Slopes and intercepts of the nine supercritical isotherms in Fig.3 obtained directly from the numerical tabulations of experimental data of the International Steam Tables [11] are collected in Table 1. These parameters can be used to estimate directly a critical temperature ($T_c$) and critical pressure ($p_c$), respectively, and also to obtain the pressures and densities, of the percolation loci PB and PA. For densities both below PB, and densities above PA, on every isotherm, the departures in pressure $\Delta p = (p-p_m)_T$, are found to be quadratic in density. A plot of density against the modulus $|(p-p_m)|^{1/2}$ gives a straight line that interpolates at $\Delta p = 0$ to obtain the percolation densities at PA and PB and, hence also the pressures using equation (2), all as given in Table 1.

The present thermodynamic description of criticality has given rise to slightly different values for the critical constants than those previously reported in the case of argon [9]. The slopes of the mesophase rigidities decrease linearly with temperature in the vicinity of $T_c$, which is then is obtainable from the experimental measurements of the rigidity constants in the mesoscale region by extrapolation of the slopes of the isotherms against temperature to zero ( Fig. 5). The present value obtained for $T_c$ by this procedure (374.66) is slightly higher than previous literature values (given as 373.746°C [11]); here the uncertainties are too great to infer revised values.

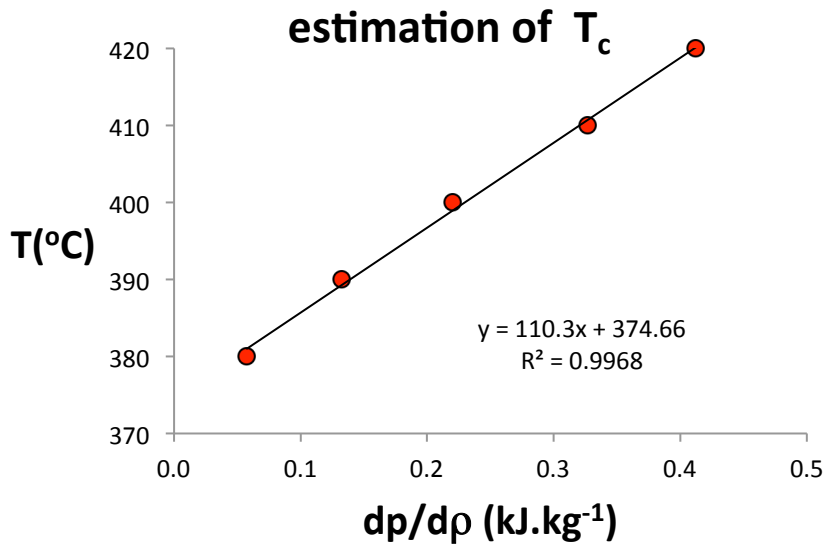

**Figure 5** A value of the critical temperature as determined from slopes of the supercritical isotherms in the meso-phase region

Likewise, a critical pressure can then also be obtained from the supercritical isotherms by plotting the intercept of the linear plots ($p_o$ in Eq. 2) against $T-T_c$, and interpolating to $T-T_c = 0$ (Fig. 6).

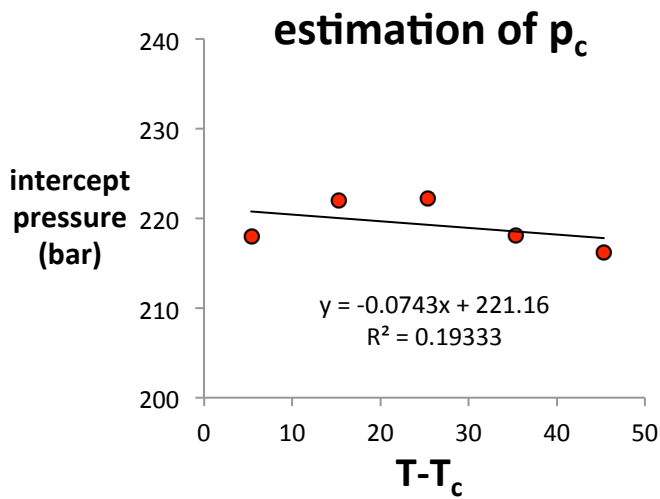

**Figure 6** Determination of a critical pressure value from the supercritical mesophase rigidity intercepts

The critical pressure we obtain by this procedure, $p_c = 221.7 \pm 1$ bar lies within the uncertainty of the literature results [11], obtained from subcritical two-phase coexistence measurements (220.64 $\pm$ 0.03 bar). Nobody has ever measured the hypothetical 'critical density' directly; ever since the critical temperature was discovered, experimentalists have used the law of rectilinear diameters (LRD) to obtain their "critical point" density [18]. This is well illustrated from the water liquid vapor coexistence densities in Fig.1. The lowest coexisting water mass density that can be measure is around 344 kg.m$^{-3}$ and the highest vapor mass density they can observe near $T_c$ is 220 kg.m$^{-3}$ as seen in Fig.1. The mean of these two experimental liquid and vapor densities is 282 kg.m$^{-3}$. This is essentially the mean value obtained from experimental saturated vapor pressures in the coexistence region, and interpolation using the LRD, historically referred to as a "critical density" [18]

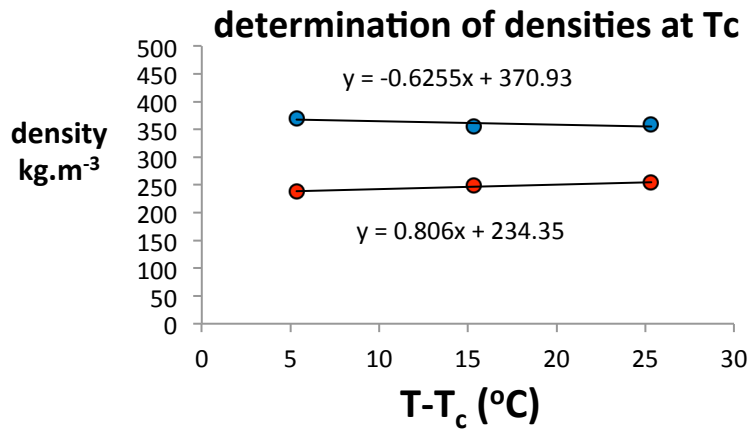

**Figure 7** Determination of the densities of coexisting water (blue circles) and steam (red circles) at the critical temperature.

The coexisting gas and liquid densities at the critical temperature and pressure can also be conveniently and accurately obtained from the pressure-dependence of the supercritical percolation loci (Table 1). The percolation densities in the range close to $T_c$ are linear functions of pressure so that the coexisting densities can be obtained by plotting the percolation densities of gas (PB) and liquid (PA) against $(p-p_c)$ and interpolating to zero as seen in Fig. 7. The uncertainties are quite large, but it is reassuring that the values obtained are not too different from the critical densities in Fig. 1.

## 5 PHASE DIAGRAM

The data for the percolation loci can now be used to construct a preliminary phase diagram of water and steam. Fig. 8 shows the p-T projection that can be compared directly with the contours of properties in Figure 2. The immediate observation is that the loci of maxima the derived properties $C_p$, $\alpha_p$ and $\kappa_T$ lie broadly in the region of the supercritical mesophase bounded by the percolation loci.

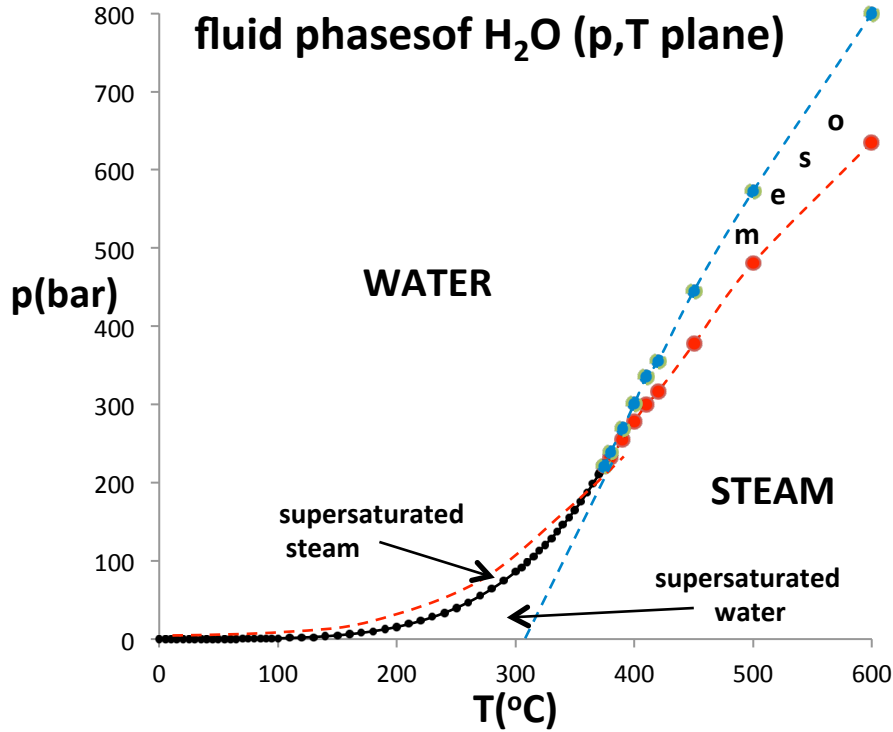

**Figure 8** Phase diagram of water in the p-T plane; the coexisting density curves are taken from the experimental measurements of the IAPWS International Steam Tables [11]; the supercritical percolation transition points PA (blue) and PB (red) intersect at $T_c$, and continue to define the matastable limit lines (usually referred to as spinodals) of the subcritical gas and liquid phases within the binodal region.

The next observation from Fig. 8 is that the critical point on the T,p plane is thermodynamically defined by an intersection of two percolation loci. All points on the Gibbs density surface must have a thermodynamic definition. The 'critical point' hypothesized by van der waals, by contrast, has no thermodynamic definition; its predicted existence was based upon a hypothesis regarding the form of the equation of state. Now, we find that two different equations of state are required for water and steam at all temperatures. Also plotted on Fig. 8 are the experimental spinodals [20] which bound the regions of metastable existence in the two-phase coexistence region at subcritical temperatures. Note also that the percolation transition loci cross at $T_c$ and continue the below $T_c$ to become the spinodals. This behavior was also observed for liquid argon [9].

At $T_c$, PA and PB cross the critical coexistence line, to become subcritical limits of existence of the metastable compressed gas ($p > p_{sat}$) and expanded liquid phases ($p < p_{sat}$), respectively. The two percolation loci define the limits of existence of gas and liquid phases, not only above $T_c$, i.e. in the supercritical meso phase, but also below $T_c$, in the two-phase water - steam coexistence region. This is consistent with a phenomenological definition of PA and PB when liquid and gas have the same values of the rigidity $(dp/d\rho)_T$ on the same isotherm whereupon $(d_2p/d\rho^2)_T = 0$ at both PA and PB, albeit at different pressures. At this point, there is also zero surface tension between water and steam, and consequently, within the mesophase, there is no barrier to spontaneous nucleation of steam from water (at PA) or water from steam (at PB).

None of the three supercritical maxima lines in Fig. 2 coincide with PA or PB although the line of maximum $\kappa_T$ on the p,T plane is quite close to PA. A better understanding of lines of maxima in these properties can perhaps be seen from the Gibbs density surface where the data points for PA and PB are plotted in Fig.9. When the various maxima in $C_p$, $\alpha_p$ and $\kappa_T$, at constant temperature, as functions of density, are obtained from the tables [11], and are put on the density phase diagram, a quite different picture emerges. All three maxima have their origins at the mean of the liquid and gas critical densities. The $C_p$ maximum, which could be Bernal's "hypercritical line" [12] essentially bisects the mesophase, at least up to 500ºC. Since $C_p$ is associated with energy fluctuations, it could represent the colloidal mesophase inversion line when the dispersed phase changes from steam to water. Curiously, both $\alpha_p$ and $\kappa_T$ also stem from the central mean density at $T_c$, and both cease even to exhibit a maximum at all nearby the points at which they intersect the two percolations lines PA and PB, respectively.

At $T_c$, the rigidity difference between liquid and gas is zero, hence the surface tension should go to zero at a finite density difference [19]. This 'critical divide' could be further understood by considering the surface tension. The percolation loci PA is the boundary of the existence of the liquid state for supercritical temperatures, it must connect up with the boundary for the non-existence of the liquid state for sub critical temperatures, i.e. the liquid-vapor decomposition spinodal. The spinodals are often defined operationally by the absence of a barrier to nucleation of the new phase, but an alternative definition is the point at which the surface tension of a liquid goes to zero as suggested by He and Attard [19].

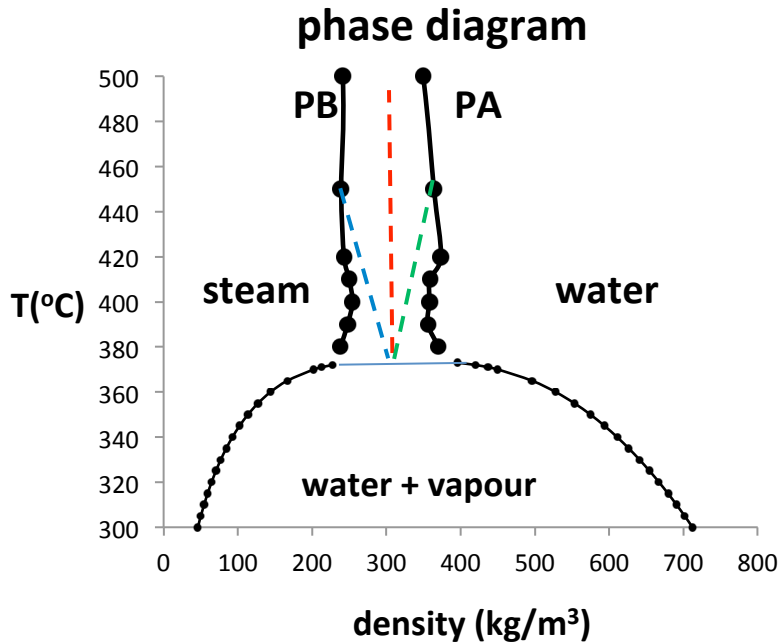

**Figure 9** Equilibrium fluid phase diagram on the Gibbs density surface; the data points for the percolation loci are given in Table 1; loci of the maxima in heat capacity ($C_p$), isothermal compressibility ($\kappa_T$) and the thermal expansivity ($\alpha_p$) (red, blue and green dashed lines respectively) along the isotherms as functions of density, from thermodynamic data source [11].

## 6 DISCUSSION AND CONCLUSIONS

In this preliminary investigation of supercritical water and steam, we have seen that the principle difference between liquid and gas is not just the density. Moreover the density difference does not disappear at the critical temperature. Rather, the structures of liquid and gas phases are fundamentally different in their description. Steam is a spatial distribution of small clusters of water molecules. Every species will have the same chemical potential with monomers being the most probable, then dimers, trimers, etc. with ever decreasing probability, to balance the decreasing entropy with the increased enthalpy of association. These hydrogen bonded molecular clusters are occupied 'sites' distributed in one large volume-spanning 'hole'. Water, on the other hand is one large volume-spanning cluster, i.e. a network, of hydrogen bonded molecular 'sites', with a distribution within it of available 'holes'. The foregoing analysis shows that this distinction between water and steam extends to supercritical temperatures.

It has been deduced from statistical thermodynamics [20] that the chemical potential is proportional to the probability of increasing the density by inserting one additional molecule into a "hole". As a consequence of this relationship the statistical properties of an average hole, relative to its environment of occupied sites and other holes, are the same as the statistical properties of the occupied site of an average molecule. We therefore expect to see a symmetry between water and steam in the work required to increase the density by simply adding one more occupied site in steam, or removing a hole, or part of a larger hole, in water by filling it with a water molecule, ensemble averaged. This is indeed seen in Fig. 10 which shows a plot of the rigidity $(dp/d\rho)_T$ i.e. the work required to increase the density at constant temperature.

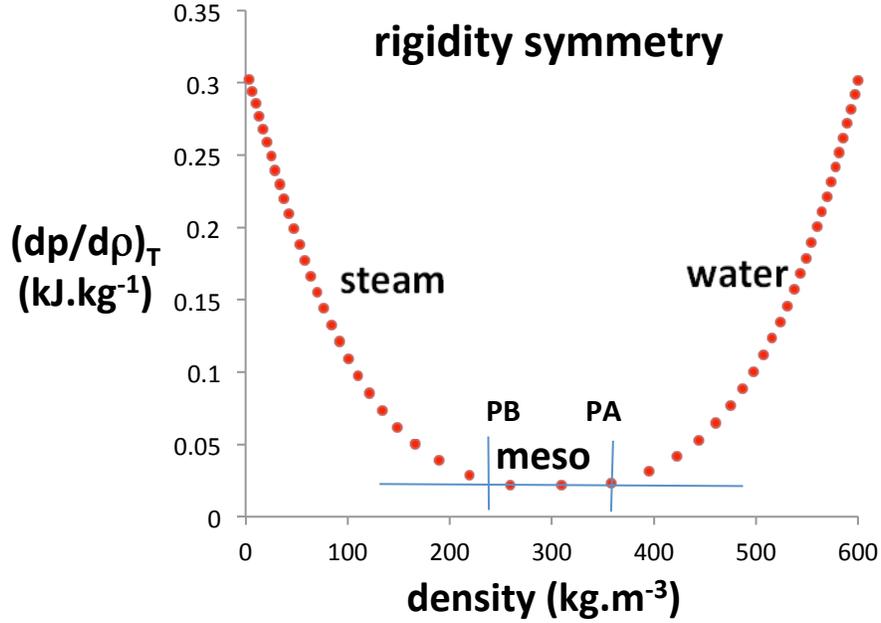

**Figure 10** Plot of $(dp/d\rho)_T$ (rigidity) for the supercritical isotherm 400°C obtained from the electronic steam tables of reference [11] showing a remarkable symmetry between steam and water, that can be explained by statistical fluctuations in density at the molecular level.

The symmetry is centered upon the mid-point of the colloidal mesophase. For the ideal gas, density zero, the function is of course analytic $(dp/dr)_T = RT$ where R is the gas constant T is in degrees Kelvin, at 400°C (673.16K) the limiting constant is 0.3111 kJ/kg. Thus, the ideal gas has a corresponding liquid state at twice the mean density between the percolation transitions PA and PB. The rigidity function continues to increase for higher liquid densities.

The relative slope of the rigidity functions is the distinguishing thermodynamic property between gases and liquids. For steam it is seen to decrease quadratically with density, and for water it increases quadratically. This can be explained by the exact statistical mechanical relationship between the rigidity function, and fluctuations in the molecular number density (N) in a thermodynamically large fixed subvolume (V) of the equilibrium fluid. Infinite snapshots of such a such a dynamical system in statistical thermodynamics is a grand canonical ensemble. From the statistical properties of this ensemble it is proven that the rigidity function is exactly related to the change in Gibbs chemical potential (μ) with number density (N) according to

$$(dp/d\rho)_T = (d\mu/d\log_e N)_{V,T} \qquad (3)$$

and from the thermodynamic theory of fluctuations [21] the right hand side is exactly and inversley proportional to an average dimensionless variance in N for a constant subvolume V, whereupon

$$(dp/d\rho)_T = RT / [<(\Delta N)^2>]_{V,T} \qquad (4)$$

where R is the gas constant, T is the absolute temperature and V is a molar volume. For the ideal gas the variance in equation (4) is 1, and it will approach zero for an amorphous ground state [7] as T→ 0K. The rigidity can never be negative; it varies from zero at the critical temperature in the mesophase, to infinity for a metastable amorphous ground state. The symmetry seen in Fig.10 is then explained by the simple observation that fluctuations in N for the occupied molecular sites in the volume of gaseous state are mirrored in a corresponding liquid state. The distribution of 'holes' in a liquid mirrors the distribution of 'sites' in the gas. Number density fluctuations in the liquid state are determined by the fluctuations in the number of 'holes'. The formal statistical theory of this translational invariance between insertion sites and holes for systems with continuous potentials was originally derived by Widom [20].

In conclusion, the previous conjecture that the present description of liquid-gas criticality, already verified in the case of argon [9], will be the same phenomenology for all liquids is here confirmed now for water and steam. The density difference between saturated water and steam does not go to zero at the critical temperature. Rather, above the critical temperature, the gas and liquid phases of water and steam are separated by a supercritical mesophase (regrettably referred to as steam II in a previous publication [8]) which is a homogeneous mixture of gas-like small clusters, and liquid-like system spanning large clusters or microdroplets, all species with the same Gibbs chemical potential in thermal equilibrium. These phase boundaries are percolation transition loci and appear here either as very weak second-order or possibly third-order thermodynamic phase transition, referred to as percolation loci. These transitions are not well understood at the molecular theory level, but now require a more refined experimental and theoretical investigation.